\documentclass[12pt,a4paper]{article}
\usepackage{amssymb}

\usepackage{graphicx}
\usepackage{amsmath}

\begin{document}

\title{Describing two-dimensional vortical flows : \\
the \emph{typhoon} case}
\author{Florin Spineanu and Madalina Vlad \\
Association Euratom-MEC Romania, NILPRP \\
MG-36, Magurele, Bucharest, Romania\\
\emph{and} \\
Research Institute for Applied Mechanics \\
Kyushu University, Kasuga 816-8580, Japan}
\maketitle

\begin{abstract}
We present results of a numerical study of the differential equation
governing the stationary states of the two-dimensional planetary atmosphere
and magnetized plasma (within the Charney Hasegawa Mima model). The most
strinking result is that the equation appears to be able to reproduce the
main features of the flow structure of a typhoon.
\end{abstract}

\section{Introduction}

There is a well known similarity between the two-dimensional models of the
planetary atmosphere and the magnetized plasma. In the absence of
dissipation the models can be reduced to differential equations having the
same structure: the Charney equation for the nonlinear Rossby waves , in the
physics of the atmosphere \cite{Charney}; and the Hasegawa-Mima equation for
drift wave turbulence, in plasma physics \cite{HM}. They are similar with
the Navier-Stokes equation because they have two conserved quantities, the
energy and the enstrophy. This in principle allows states of negative
temperatures, or, equivalently, these models support a trend to organised
vortical flow. It results the possibility to have as solutions coherent
structures (vortices) besides the turbulent states characterised by spectral
cascade.

These analytical models have led to a serious advancement of our knowledge
in both fields. However the stationary states appear to be described within
these models by a reduced equation having a too wide generality,
representing actually something as a constraint with weak ability to
identify unequivocally the real solutions: it simply states that at
stationarity the advection of the vorticity by the velocity vector field
vanishes. In reality, numerical simulations show that the stationary states
reached in relaxation are very regular and persist for a long time period
and that this set of asymptotic states is not the huge space of functions
able to fulfill the constrained mentioned above. The fluid evolves at
relaxation toward a reduced subset of functions, characterized by regular
shape of the streamfunction \cite{HTK}, \cite{KMcWT}, \cite{KTMcWP}, \cite
{HH} (and references therein). At the oposite limit the turbulent regime can
be treated with renormalization group methods \cite{DiamondKim}.

It is well-known that the same phenomenon exists in the case of the ideal
fluid described by the Euler equation. By experiments and numerical
simulation it has been shown that the ideal fluid evolves at relaxation
toward a very ordered flow pattern, consisting of two (positive and
negative) vortices and that this state persists for very long times, being
limitted by only the effect of some residual dissipation. From numerical
simulations it has also been inferred the form of the flow function. It has
been found that the streamfunction obeys, in these states, the \emph{sinh}%
-Poisson equation. Montgomery and his collaborators have developed a
theoretical statistical model which explains the appearence of this equation
in this context \cite{Montgomery1}, \cite{Montgomery2}, \cite
{KraichnanMontgomery}, \cite{Montg2}, \cite{Montg3}, \cite{Joyce}, \cite
{Smith}. Later, the equation has also been derived by formulating the
continuum version of point-like vortices as a field theoretical model of
interacting gauge and matter fields in the adjoint representation of $%
SU\left( 2\right) $ \cite{FlorinMadi1}. The essential point of the latter
derivation was the self-duality of the relaxation states of the fluid.

No equation (similar to the \emph{sinh}-Poisson equation in the Euler fluid
case) has been found for the Charney-Hasegawa-Mima (CHM) equation, despite a
considerable effort \cite{Seyler}, \cite{Montg4}. However, as mentioned
before, there are convincing experimental and numerical indications that the
fluids (atmosphere and plasma) evolve to a reduced subset of states.

\bigskip

We have developed a field theoretical model for the point-like vortices with
short range interaction, based on Chern-Simons action for the gauge field in
interaction with the nonlinear matter field, again in $SU\left( 2\right) $
algebra. It is then possible to derive the energy as a functional that
becomes extremum on a subset of stationary states and presents particular
properties. The general characterization of this family of states is their 
\emph{self-duality}, which here means that the energy functional becomes
minimum because the square terms are all vanishing, leaving as lower bound a
quantity with topological meaning. A very detailed account of the derivation
is in Refs. \cite{FlorinMadi2}, \cite{Toki2003}.

The result is a set of equations parametrized by the solutions of the
Laplacean equation in two-dimensions.

The simplest of these equations is 
\begin{equation}
\Delta \psi +\frac{1}{2p^{2}}\sinh \psi \left( \cosh \psi -p\right) =0
\label{eq}
\end{equation}
(where $p$ is a positive constant). There are already some confirmations
that this is the equation governing the asymptotic stationary states of the
CHM fluids : the scatterplots of $\left( \psi ,\omega \right) $ =
(streamfunction, vorticity) obtained in experiments \cite{expgeo} and the
scatterplots obtained in numerical simulations \cite{Seyler} are very
similar to the nonlinear term of Eq.(\ref{eq}).

The objective of this work is to provide the first elements resulting from a
numerical investigation of this equation.

The results are summarised here. This differential equation is able to
reproduce the main two-dimensional features of the typhoon vortical flow. In
the physics of the atmosphere, it seems that other examples, like the
tropical cyclones, can be reproduced by solutions of this equation. The
following are the features we consider as very particular to the typhoon
morphology (in $2D$) \ \cite{Andrew}, \cite{cycrev}, \cite{ReMont}, \cite
{mesov}:

\begin{enumerate}
\item  The very narrow dip of the azimuthal velocity (mean tangential wind)
in the center of the vortex, compared with the very large extension in
space. This is characterized by the ``radius of the maximum tangential
wind'' and this radius, as mentioned, is much smaller than the diameter of
the vortex. Our equation is able to generate solutions with this structure.

\item  The slow decay of the magnitude of the azimuthal velocity toward the
periphery, compared with the very fast decay toward the center; this is
reproduced by the solutions of this equation.

\item  The very low magnitude (almost vanishing) of the vorticity over most
of the vortex (approx. from the radius of maximum wind to the periphery),
while the magnitude in a narrow central region is extremely high. This
feature is also reproduced by the equation.

\item  quantitatively, we obtain for the diameter of the typhoon's eye a
relatively good magnitude. The vorticity is higher than in observations but
not far from the realistic range.
\end{enumerate}

\bigskip

We have very encouraging results of studies on plasma vortices, but they are
not reported here. In plasma physics, the symmetrical, stable, vortical
structures observed in experiments in the linear machine seem to belong to
the class of solutions of this equation. We have also obtained several
solutions that are very similar to the crystals of vortices, known from
experiments.

\section{Numerical studies of the equation}

The numerical solution of this equation appears to be very difficult. This
may be explained by the fact that the exponentials of the two functions $%
\sinh $ and $\cosh $ are very rapidly-varying functions and any perturbation
is amplified and propagated in the solution.

In addition, the Laplace operator has spurious solutions with exponential
behavior that have to be eliminated by the numerical procedure.

The paper of McDonald \cite{McDonald} on the numerical integration of the 
\emph{sinh}-Poisson equation is very helpful in understanding the problems
related to a numerical treatment of our equation. However the approach
proposed in that paper requires to use a small mesh, specifically for
excluding the spurious modes of the Laplacean. In the case of our equation,
the vortices require a reasonable detailed description and this needs larger
meshes. Then the problem of the precision of integration procedure arises
and, if the initialization happens to be far from one of the solution, the
number of iteration of the solver is high and the errors accumulate, leading
to lack of convergence. It may be supposed that the solutions would be
similar to those of the \emph{sinh}-Poisson equation, but structures with
sharp spatial variation may be possible \cite{MontgPriv}.

\bigskip

The structure of the function space representing the union of attractors for
the various solutions of this equation appears to be very complex. This
immediately translates into serious obstacles in the attempt to reach one of
the presumed solution. The main instrument is, naturally, the
initialization, \emph{i.e.} to start the integration in the right subspace,
representing the attractor of that solution. Since there is no available
analytical description of this space, the search is simply a problem of
guessing a reasonable initial function and to repeat as many times as
necessary. One of the specific behaviors is the tendency of driving the
solution toward the constant value 
\begin{equation}
\psi =\psi _{b}^{\left( 1,2\right) }  \label{psi12b}
\end{equation}
(see Eq.(\ref{bcon})) which trivially verifies the equation. This seems to
imply that there is a large attractor in the function space around these
constant solutions. The solution which is larger in absolute magnitude is
less stable since any fluctuation around the constant generates high
vorticity. We underline that the integrations described here are \textbf{not}
radial (\emph{i.e.} unidimensional).

With all the difficulties of getting a right initial positioning in the
integration procedure we note however that the solution with the \emph{%
typhoon} morphology appears instistently from a wider class of initial
shapes.

\subsection{The numerical code}

We use the code ``\textbf{GIANT} A software package for the numerical
solution of very large systems of highly nonlinear systems'' written by U.
Nowak and L. Weimann \cite{giant}. The code belongs to the numerical
software library \emph{CodeLib} of the \textbf{Konrad Zuse Zentrum fur
Informationstechnik Berlin}. The meaning of the abbreviation is: GIANT =
Global Inexact Affine Invariant Newton Techniques and corresponds to the
implementation of the method proposed by Deuflhard (for many references see 
\cite{giant}).

This code solves nonlinear problems 
\begin{eqnarray}
F\left( x\right) &=&0  \label{gian} \\
\text{initial guess of solution, }x &=&x_{0}  \notag
\end{eqnarray}
The global affine invariant Newton schemes requires the solution of linear
problems. For higher accuracy meshes the linear problems are solved by
iterative methods. The balance between numerical requirements of the Newton
iteration (called \emph{outer} iteration) and the iterative linear solver (%
\emph{inner}) means that the solution of the linear problem will be
approximative. Two packages of linear solvers can be used, GMRES
(generalized minimum residual : Brown, Hindmarsh, Seager) and GBIT1 (fast
secant method using the \emph{Good-Broyden} updates : Deuflhard, Freund and
Walter).

All necessary description of the method, of the code and many studies of the
numerical precision and computer efficiency are presented by Nowak and
Weimann in the documentation of the code.

The code has been implemented and the tests have been performed with
successful results (we are grateful to Dr. Weimann for his kind help in this
problem).

\subsection{Boundary conditions}

The boundary conditions are dependent on the value of $p$. The physical
model imposes that the scalar function $\psi $ remains nonzero at infinity
for $p>1$. This means that we must require that the boundary condition is
one of the roots of the algebraic equation 
\begin{equation}
\cosh \psi -p=0  \label{coshp}
\end{equation}
which can give the vanishing of the physical vorticity at infinity. Then we
impose 
\begin{eqnarray}
\text{boundary condition }\psi \left( r\rightarrow \infty \right) &=&\psi
_{b}^{\left( 1,2\right) }  \label{bcon} \\
&=&\ln \left( p\pm \sqrt{p^{2}-1}\right)  \notag
\end{eqnarray}

\subsection{Initialization}

In general the initial profiles has been of two types: symmetric profiles
with maximum centered on $\left( 0,0\right) $ and initializations with
functions expressed as product of trigonometric functions.

The symmetric profiles has been chosen as Gaussian functions, or various
annular shapes.

For may runs, as suggested by the experiments for the \emph{sinh}-Poisson
equation (paper by McDonald \cite{McDonald}), the initial function is taken
as a product of trigonometric functions in both directions, $x$ and $y$. We
need to prepare the initial function in the sense that the values that are
obtained in for the vorticity, \emph{i.e.} the Laplacean of the initial
distribution should not be too different of what is obtained by simply
inserting the initial function in the nonlinear term. For this we take a
coefficient $\psi _{in}$ of the product of the trigonometric functions as a
parameter to be determined.

The initial function is taken as 
\begin{equation}
\psi \left( x,y\right) =\psi _{b}^{\left( 1\right) }+\psi _{in}\sin \left(
k\pi \frac{x-x_{\min }}{x_{\max }-x_{\min }}\right) \sin \left( k\pi \frac{%
y-y_{\min }}{y_{\max }-y_{\min }}\right)  \label{trig}
\end{equation}
where $k$ is the periodicity of the profile and $\psi _{in}$ is the
amplitude. We insert in the equation and we require approximative equality
of the two parts, the vorticity and the nonlinearity. This is obtained by
choosing a point $\left( x,y\right) $ where the initial function is maximum
and it results a condition on only the amplitude, $\psi _{in}$. 
\begin{eqnarray}
\Delta \psi &=&\psi _{in}\left[ 2\left( k\pi \right) ^{2}\right]
\label{gues} \\
&\simeq &\frac{1}{2p^{2}}\sinh \psi _{in}\left( \cosh \psi _{in}-p\right) 
\notag
\end{eqnarray}
This equation is solved and one of the roots is selected as the amplitude of
the initial function.

\bigskip

The experiments with simple $\sin $ functions frequently lead to
difficulties of convergence. Looking at the function's form (either partial
evolutions during iterations or good, converged, results) we notice that the
two-signed values are less tolerated and only one of the signs survives.
This led us to adopt forms expressed as square of the trigonometric
functions.

\section{Results of the numerical integration}

\subsection{The typhoon morphology}

The value of the parameter is $p=1$. The domain is 
\begin{equation*}
\left( x,y\right) \in \left[ -0.5,0.5\right] \times \left[ -0.5,0.5\right]
\end{equation*}
with\ $\left[ 101,101\right] $ mesh points. The boundary value is 
\begin{equation*}
\psi _{b}^{\left( 1\right) }=\ln \left( p-\sqrt{p^{2}-1}\right) =0
\end{equation*}
and the initial function is 
\begin{equation*}
\psi \left( x,y\right) =\psi _{b}^{\left( 1\right) }+4.2\times \sin \left(
4\pi \frac{x-x_{\min }}{x_{\max }-x_{\min }}\right) \sin \left( 4\pi \frac{%
y-y_{\min }}{y_{\max }-y_{\min }}\right)
\end{equation*}
It takes $501$ calls to the function and Jacobian. The accuracy is $%
0.257\times 10^{-3}$. This run has been executed with several mesh
dimensions: $\left[ 31\times 31\right] $, $\left[ 51\times 51\right] $, $%
\left[ 71\times 71\right] $. The results are very close, but higher accuracy
shows much clearer the details.

\begin{figure}[tbph]
\centerline{\includegraphics[height=5cm]{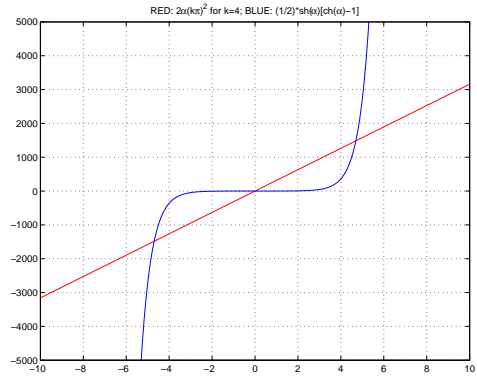}}
\caption{The procedure to find an approximation to a good initialization.}
\label{alfa_k4_7}
\end{figure}

\begin{figure}[tbph]
\centerline{\includegraphics[height=5cm]{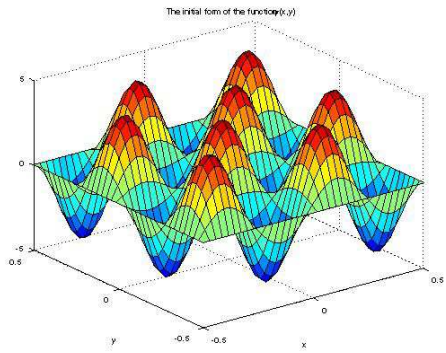}}
\caption{The initial function, trigonometric profiles.}
\label{exp_7b}
\end{figure}

The results are shown. The Figure (\ref{alfa_k4_7}) shows the choice of the
amplitude of the initialization and Fig.(\ref{exp_7b}) shows the initial
function $\psi $.

\begin{figure}[tbph]
\centerline{\includegraphics[height=5cm]{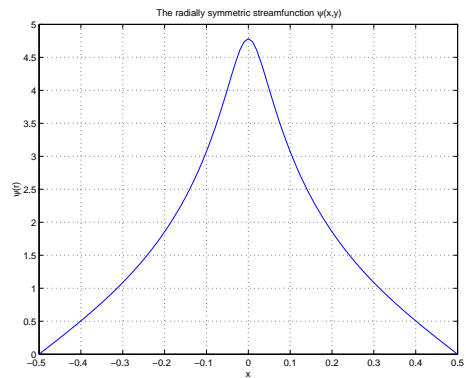}}
\caption{The section along $x$ of the solution $\protect\psi(x,y)$.}
\label{exp_7c}
\end{figure}

\begin{figure}[tbph]
\centerline{\includegraphics[height=5cm]{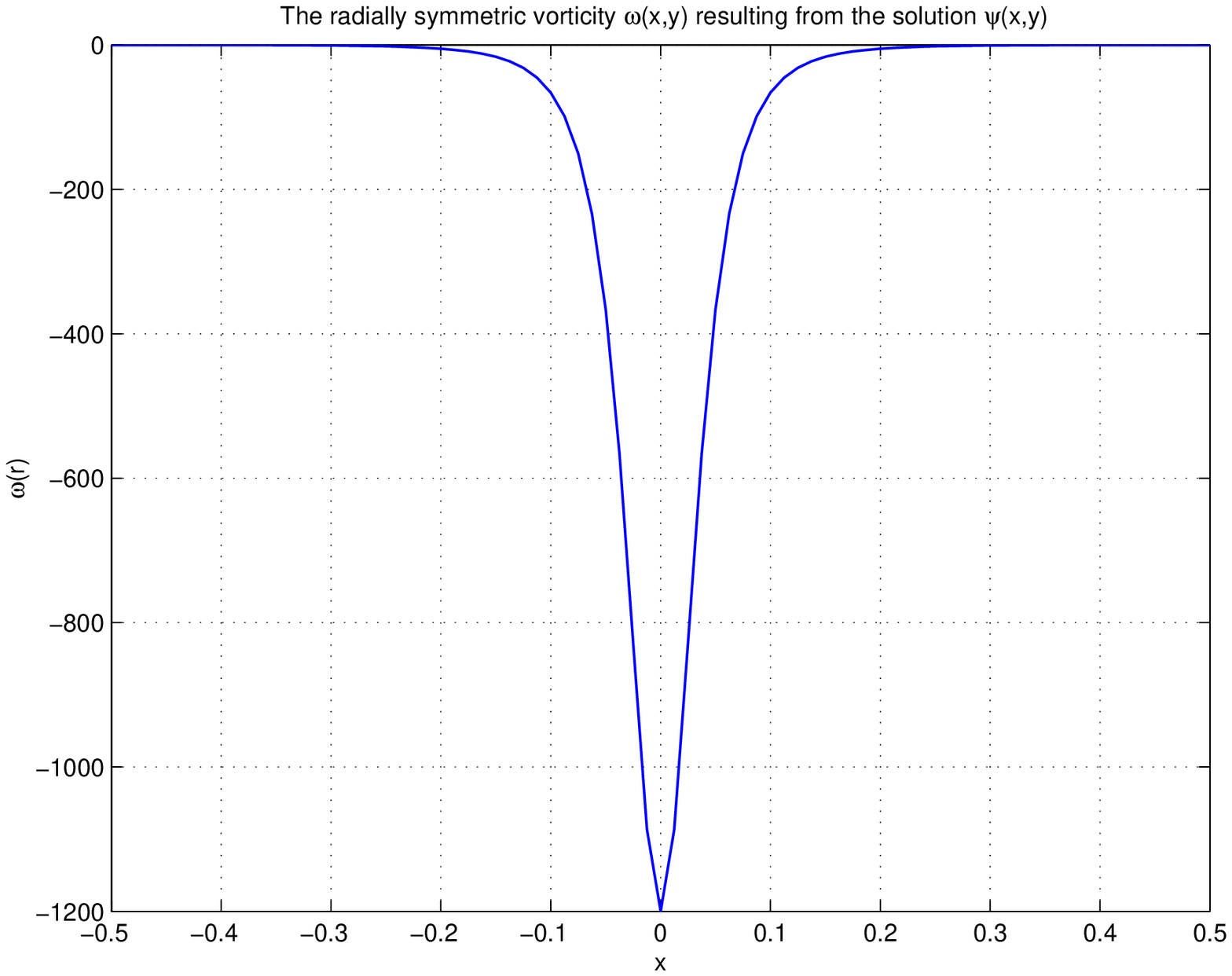}}
\caption{The vorticity, calculated from $\protect\psi (x,y)$ obtained by
integration.}
\label{omega_7}
\end{figure}

The solution has an apparent cylindrical symmetry around the center and for
this reason we present a section along $x$ of the streamfunction $\psi
\left( x,y\right) $ (Fig.(\ref{exp_7c})). A section along $x$ axis of the
vorticity $\omega \left( x,y\right) $ is presented in Fig.(\ref{omega_7}).


\begin{figure}[tbph]
\centerline{\includegraphics[height=5cm]{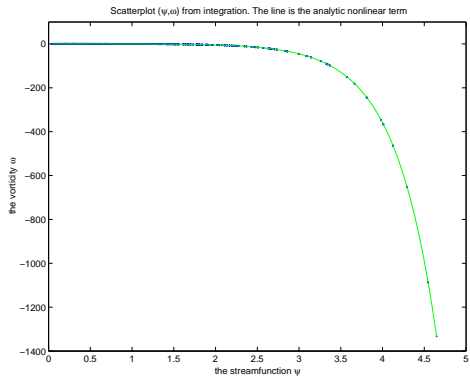}}
\caption{The scatter plot $(\protect\psi ,\protect\omega )$, for $p=1$.}
\label{comparatie_A_7}
\end{figure}

\begin{figure}[tbph]
\centerline{\includegraphics[height=5cm]{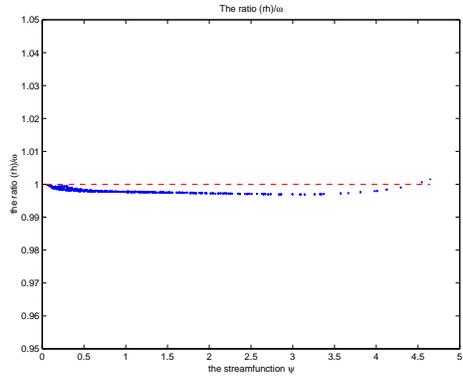}}
\caption{The ratio of $\protect\omega $ and the nonlinear term.}
\label{comparatie_C_7}
\end{figure}

In order to quantify the accuracy of integration we collect in all the
domain $\left( x,y\right) $ the pairs $\left( \psi ,\omega \right) $ and
represent them together with the line of the nonlinear term in our equation,
Fig.(\ref{comparatie_A_7}). In Fig.(\ref{comparatie_C_7}) we show the ratio
of the two quantities the nonlinear term and $\omega $, as resulted from the
calculated $\psi $. This ratio should be $1$. There are points close to the
value $0$ where this ratio is not $1$ but, if we normalize adding an
arbitrary constant to remove the possible singular cases, we notice a very
good clustering of the points around the line $1$. In addition, we represent
the scatterplot of the pairs ($\omega $, magnitudes of nonlinear term for
the $\psi $'s) and notice the close clustering around the diagonal. Other
tests are possible and they indicates that the integration is very good on
most of the region and good within the imposed accuracy in the regions where
the quantities reach values close to $0$.

\begin{figure}[tbph]
\centerline{\includegraphics[height=5cm]{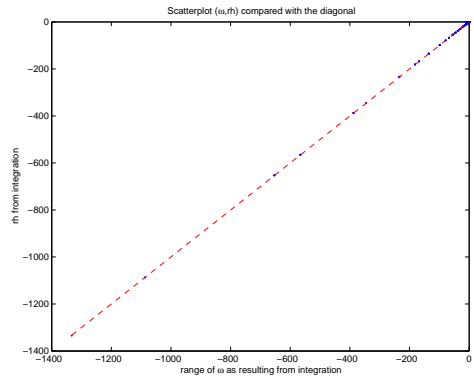}}
\caption{Scatterplot $(\protect\omega ,$ the nonlinear term), compared with
the diagonal.}
\label{comparatie_E_7}
\end{figure}

\begin{figure}[tbph]
\centerline{\includegraphics[height=10cm]{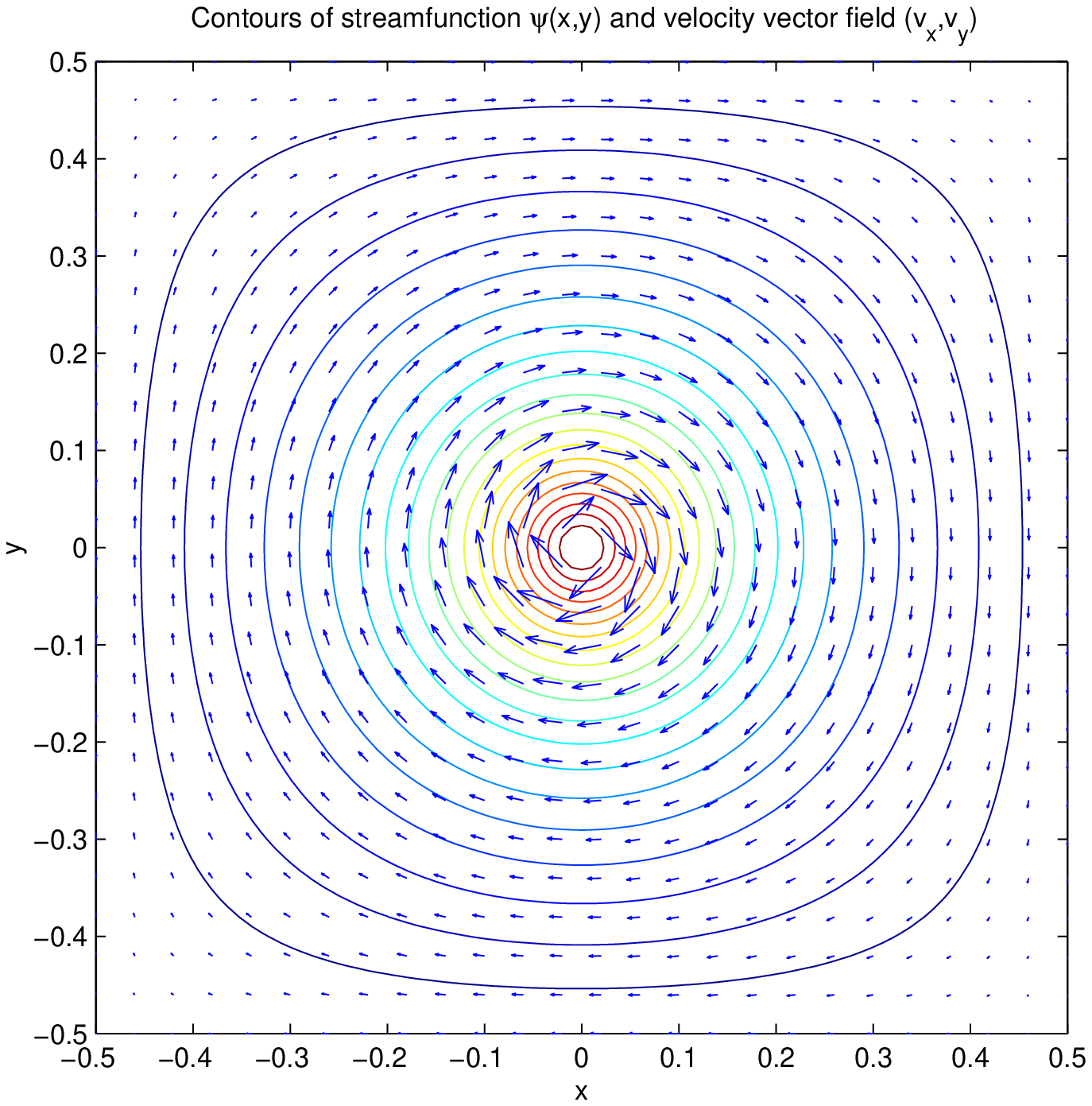}}
\caption{The contours of the scalar streamfunction $\protect\psi (x,y)$ and
the vector field $(v_{x}.v_{y})$.}
\label{vel_7}
\end{figure}

The contour plot of the solution is shown in Fig.(\ref{vel_7}) on the same
graph with the velocity field (we have used a reduced set of data due to
limitations on the EPS file). We must note that this two-dimensional
integration gives a radial component of the velocity which at maximum is
about $20$ times lower than the tangential one.

\begin{figure}[tbph]
\centerline{\includegraphics[height=10cm]{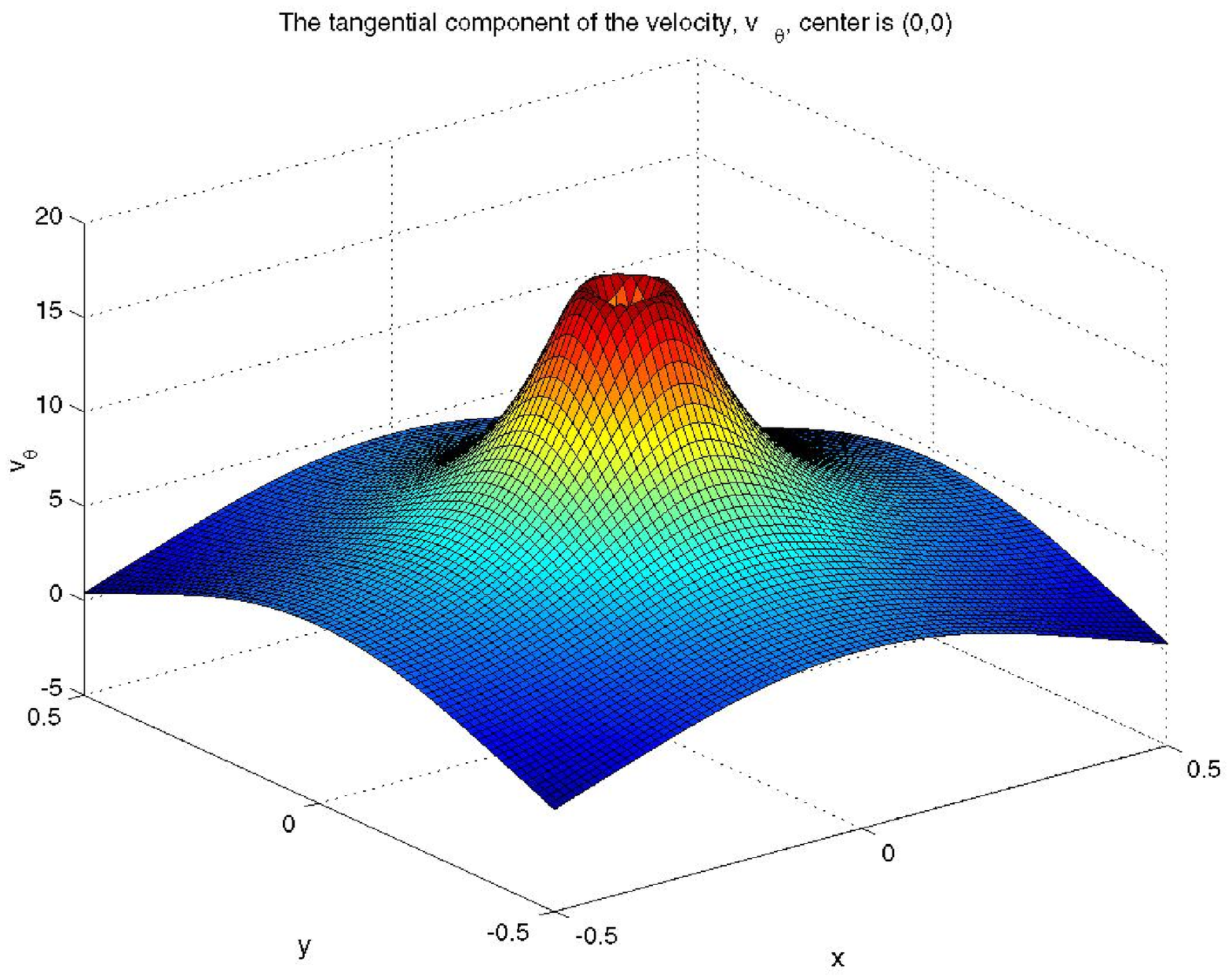}}
\caption{The tangential component $v_{\protect\theta }(x,y)$ of the velocity
vector field $(v_{x}.v_{y})$, with center at $(0,0)$.}
\label{vth_7a}
\end{figure}

The tangential component of the velocity is shown in two figures (\ref
{vth_7a}) and (\ref{vth_7c}) with the purpose of making easier the
observation of the central region. The narrow dip in the center is clearly
visible and its radial extension can be compared with the extension of the
whole domain.

We have represented in Fig.(\ref{vthlin_7}) a section along the $x$ axis of
the amplitude of the azimuthal component of the velocity.

\begin{figure}[tbph]
\centerline{\includegraphics[height=10cm]{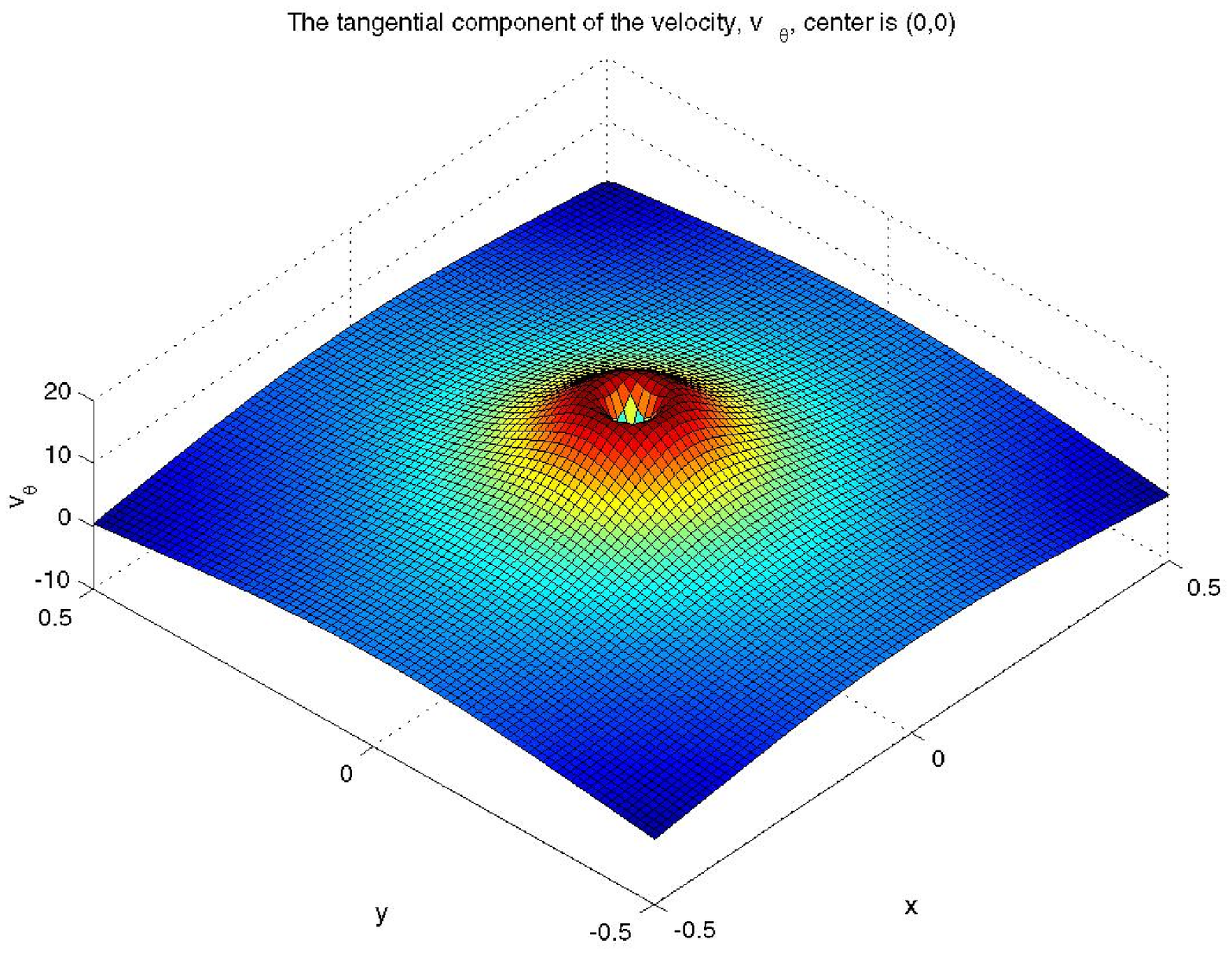}}
\caption{The tangential component $v_{\protect\theta }(x,y)$ of the velocity
vector field $(v_{x}.v_{y})$, with center at $(0,0)$ (same as \ref{vth_7a}).}
\label{vth_7b}
\end{figure}
\begin{figure}[tbph]
\centerline{\includegraphics[height=10cm]{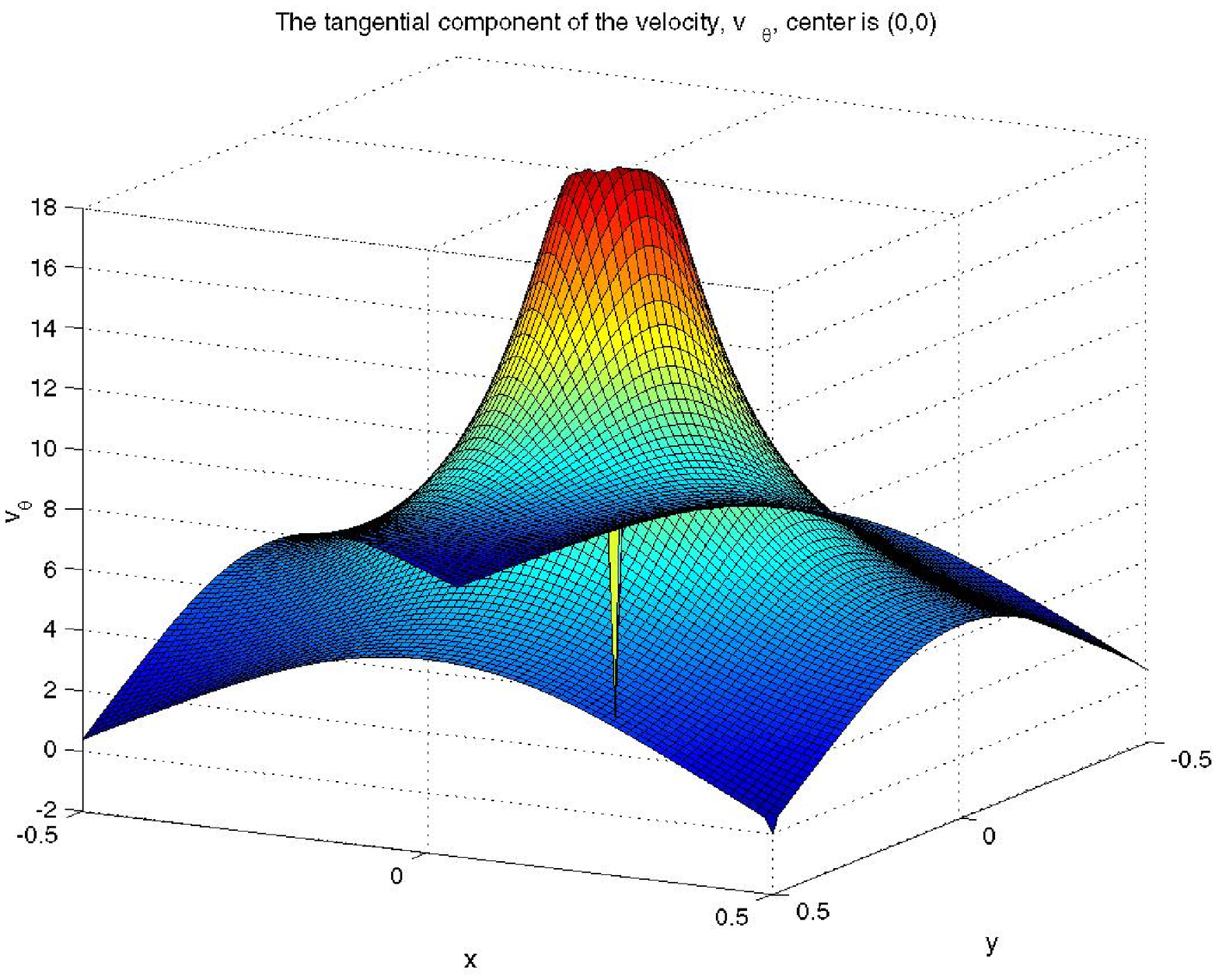}}
\caption{The tangential component $v_{\protect\theta }(x,y)$ of the velocity
vector field $(v_{x}.v_{y})$, with center at $(0,0)$ (same as \ref{vth_7a}).}
\label{vth_7c}
\end{figure}
\begin{figure}[tbph]
\centerline{\includegraphics[height=10cm]{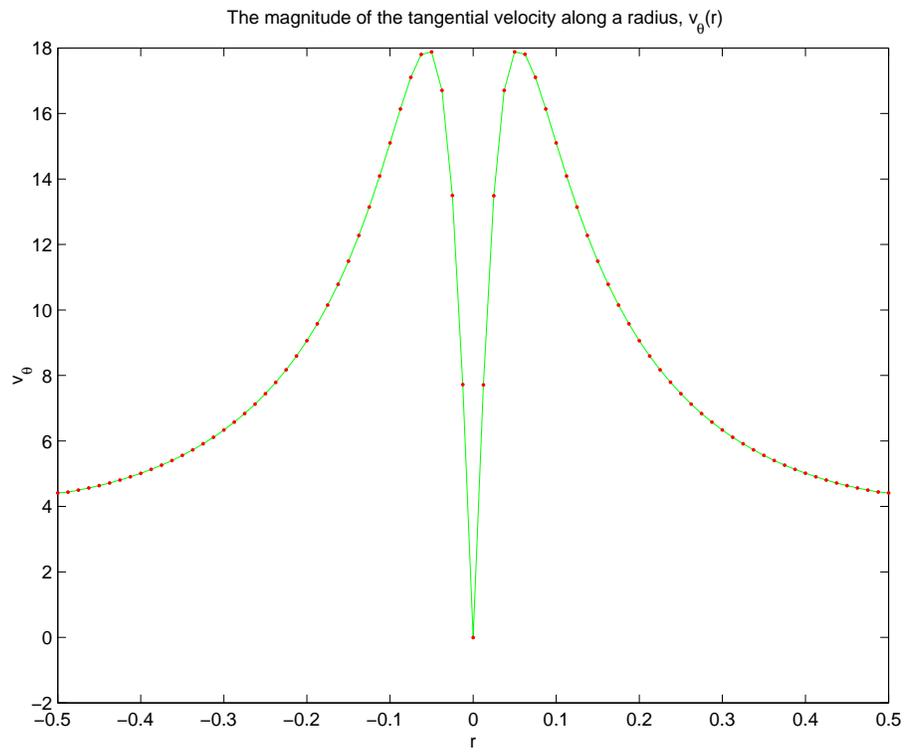}}
\caption{The magnitude of the tangential component $v_{\protect\theta }(x,y)$%
, seen along a radial line. The central fast decay is clearly visible.}
\label{vthlin_7}
\end{figure}

\subsubsection{Episodic structure of two vortices}

\begin{figure}[tbph]
\centerline{\includegraphics[height=5cm]{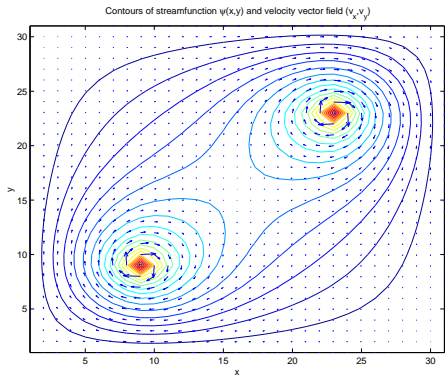}}
\caption{The contours of the scalar streamfunction $\protect\psi (x,y)$ and
the vector field $(v_{x}.v_{y}$ for a two-vortices approximative solution.}
\label{vel_4}
\end{figure}

It is worth to mention that in a numerical experiment we have identified a
state where two vortices have been formed, placed in symmetrical positions
along the diagonal of the square domain $\left[ -0.5,0.5\right] \times \left[
-0.5,0.5\right] \;$with a mesh of\ $\left[ 31,31\right] $. The value of the
parameter is $p=1$. The initial function is trigonometric with $k=2$ in Eq.(%
\ref{trig}) with a coefficient $\psi _{lin}=3.8$. It takes longer to obtain
the solution with $0.84\times 10^{-4}$ accuracy, $389$ calls to the
function. The result is in Fig.(\ref{vel_4}).

This state has been reexamined with much higher accuracy. It has taken long
time to see that the final solution was again the centered vortex shown
before. Therefore from the point of view of the numerical experience this
state of two vortices is irrelevant. However, the persistence of this state
inside the iterative search may indicate that it is close to a solution,
possible less stable. We have not investigated this further. Instead we will
show below a solution with four vortices.


\subsubsection{Four vortices}

The calculations are done for $p=1$ on meshes with various levels of
details: $31$, $61$, $101$. The initial function is trigonometric with $k=3$.

The results show clearly the formation of four vortices, as shown by Fig.(%
\ref{vel_3}). Each of them has a structure that is similar to the one
presented in Fig.(\ref{exp_7c}). It is interesting to note that again the
vorticity is almost zero everywhere on the domain, except a strict region
around the four vortices, where it reaches very high values.

To the accuracy we have used unitl now we cannot say if the local tangential
velocity presents the same very fast decay to the center of the vortex.

\begin{figure}[tbph]
\centerline{\includegraphics[height=10cm]{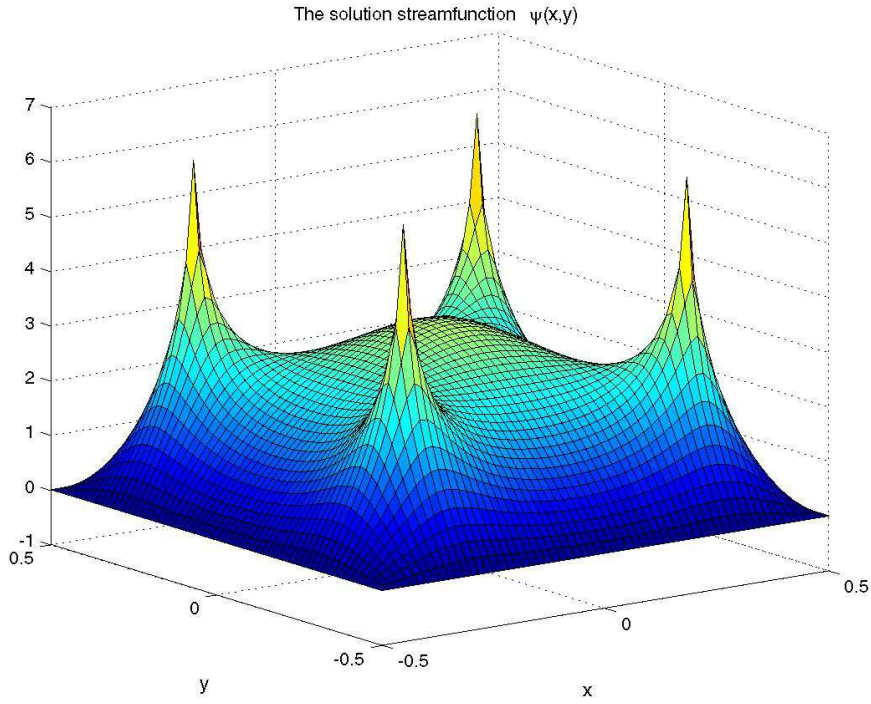}}
\caption{The scalar streamfunction $\protect\psi (x,y)$ for a four-vortices
solution.}
\label{solution_3}
\end{figure}
\begin{figure}[tbph]
\centerline{\includegraphics[height=10cm]{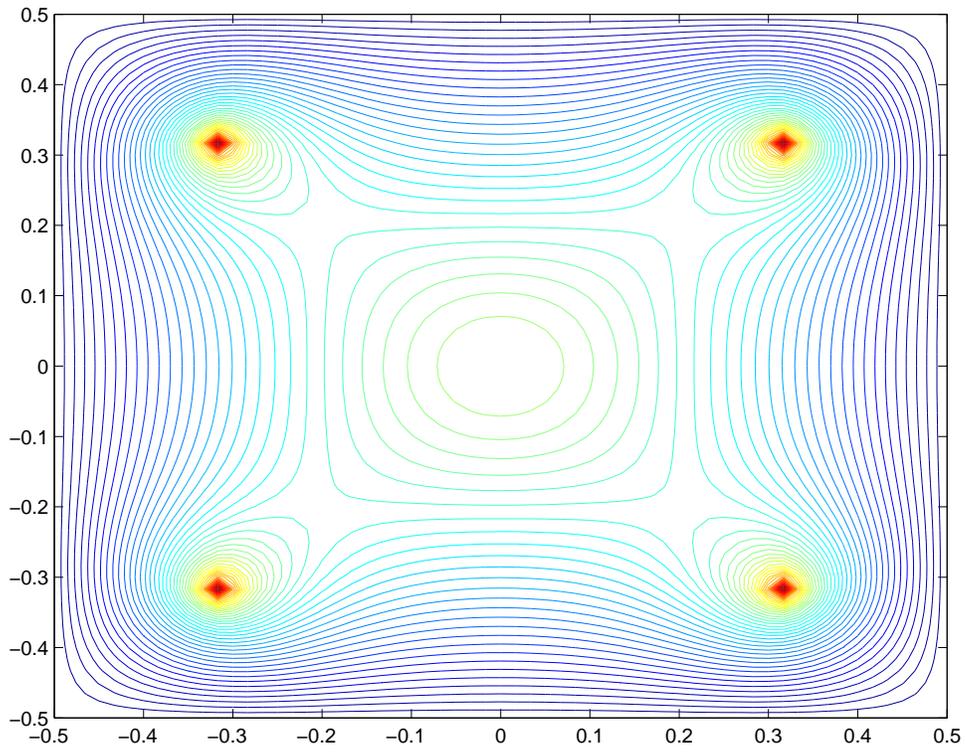}}
\caption{The contours of the scalar streamfunction $\protect\psi (x,y)$ and
the vector field $(v_{x}.v_{y})$ for a four-vortices solution.}
\label{vel_3}
\end{figure}

\subsubsection{Four vortices obtained at $p>1$}

For $p=3$ it is also possible to obtain the four-vortex solution. The
initial function is here a trigonometric combination for, $k=2$ and squared
such that only positive (four) maxima are initially present, with an
amplitude of about $\psi = 4$.


\subsubsection{The central strong decay of the tangential velocity, at $p>1$}

The numerical integration is done for $p=3$ , using an initialization by a
centered peak from an trigonometric function.

\begin{figure}[tbph]
\centerline{\includegraphics[height=10cm]{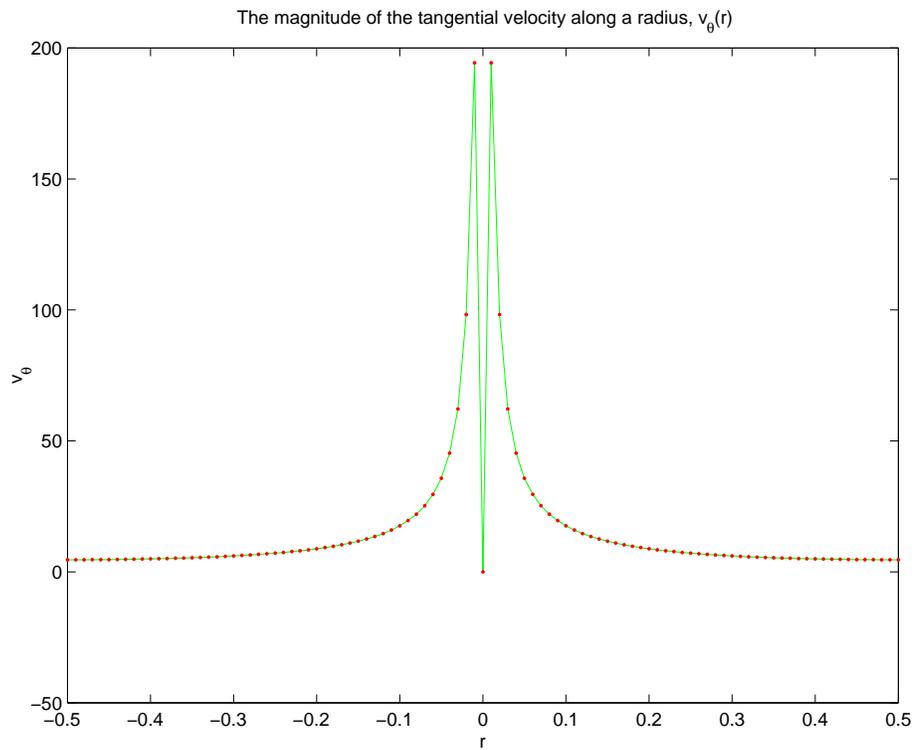}}
\caption{The magnitude of the tangential component $v_{\protect\theta }(x,y)$%
, seen along a radial line. The central dip is visible but significantly
narrower than at $p=1$.}
\label{vthlin_10}
\end{figure}

We note from Fig.(\ref{vthlin_10}) that for larger values of the parameter $%
p $ there is a even more narrow zone where there is the strong decay of the
tangential velocity.


\subsection{Relevance of the solutions for the physics of the atmosphere}

In general the space variables of the CHM equation are normalized to the
intrinsic typical length of the model. In this case (atmospheric physics)
are scaled with $\rho _{g}$, the Rossby radius. We note in passing,
(especially for plasma physicists) that there is a major difference compared
with the plasma case. In plasma, perturbations with lengths less or
comparable with an ion Larmor gyroradius $k^{-1}\gtrsim \rho _{i}$ cannot be
described by fluid models.

In the physics of atmosphere, the wavelengths can be much smaller 
\begin{equation*}
k\rho _{g}\gg 1
\end{equation*}
At very large $k\rho _{g}$ the description becomes governed by the Euler
equation (see \cite{HH}).

For the numerical studies we choose 
\begin{eqnarray*}
\left( x,y\right) &\in &\left[ x_{\min },x_{\max }\right] \times \left[
y_{\min },y_{\max }\right] \\
&=&\left[ -0.5,0.5\right] \times \left[ -0.5,0.5\right]
\end{eqnarray*}
This means that the full domain (the side of the rectangle) is a single unit
length $\rho _{g}$.

\bigskip

In the following we make few consideration about what we can expect as
results, in the case of the atmosphere problem.

As we will notice from numerical solution, the equation produces functions
with very clear similarity with the \emph{typhoon} morphology. The
characteristic aspect is (within the precision of these first integrations)
a sharp extremum of the vorticity on $\left( 0,0\right) $ which means a
localised maximum of the tangential velocity $v_{\theta }$ in close
proximity of the center. Since 
\begin{equation*}
v_{\theta }=\frac{d\psi }{dr}
\end{equation*}
the maximum at 
\begin{equation*}
r=a
\end{equation*}
means 
\begin{equation*}
\frac{dv_{\theta }}{dr}=\frac{d^{2}\psi }{dr^{2}}=0
\end{equation*}
The equation is 
\begin{equation*}
\frac{d^{2}\psi }{dr^{2}}+\frac{1}{r}\frac{d\psi }{dr}=\left( -\frac{1}{%
2p^{2}}\right) \sinh \psi \left( \cosh \psi -p\right)
\end{equation*}
We multiply by $r\,\ $and we make a derivation to $r$%
\begin{eqnarray*}
&&\frac{d^{2}\psi }{dr^{2}}+r\frac{d^{2}}{dr^{2}}\frac{d\psi }{dr}+\frac{%
d^{2}\psi }{dr^{2}} \\
&=&\left( -\frac{1}{2p^{2}}\right) \left\{ \sinh \psi \left( \cosh \psi
-p\right) +\right. \\
&&\left. +r\left( \frac{d\psi }{dr}\right) \left[ \cosh ^{2}\psi -p\cosh
\psi +\sinh ^{2}\psi \right] \right\}
\end{eqnarray*}
We calculate this expression and the equation in the point $r=a$ defined as
the point of the maximum of the tangential velocity. This means 
\begin{eqnarray*}
r &=&a \\
\left( \frac{d^{2}\psi }{dr^{2}}\right) _{a} &=&0 \\
\left. \frac{d^{2}}{dr^{2}}\left( \frac{d\psi }{dr}\right) \right| _{a}
&\equiv &\left. \frac{d^{2}v_{\theta }}{dr^{2}}\right| _{a}=-\alpha \;\text{%
where\ }\alpha >0 \\
\psi \left( r=a\right) &\equiv &\psi _{0}
\end{eqnarray*}
where we have introduced a notation for the value, $-\alpha <0$ of the
second derivative of the tangential velocity at its maximum. For a \emph{%
very qualitative} estimation, used in predicting shapes of solutions, we
will take this as a parameter. At the point $r=a$ the equation becomes 
\begin{equation*}
\frac{1}{a}\left( \frac{d\psi }{dr}\right) _{a}=\left( -\frac{1}{2p^{2}}%
\right) \sinh \psi _{0}\left( \cosh \psi _{0}-p\right)
\end{equation*}
In the equation derivated at $r$ we replace $d\psi /dr$ with its value from
the above equation and also introduce the parameter $\alpha $. Then we have 
\begin{eqnarray*}
&&a\left( -\alpha \right) \\
&=&\left( -\frac{1}{2p^{2}}\right) \left\{ \sinh \psi _{0}\left( \cosh \psi
_{0}-p\right) \right. \\
&&+a^{2}\left( -\frac{1}{2p^{2}}\right) \sinh \psi _{0}\left( \cosh \psi
_{0}-p\right) \\
&&\left. \times \left( 2\cosh ^{2}\psi _{0}-p\cosh \psi _{0}-1\right)
\right\}
\end{eqnarray*}
or 
\begin{equation*}
\frac{a\alpha \left( 2p^{2}\right) }{\sinh \psi _{0}\left( \cosh \psi
_{0}-p\right) }=1-\frac{a^{2}}{2p^{2}}\left( 2\cosh ^{2}\psi _{0}-p\cosh
\psi _{0}-1\right)
\end{equation*}
This equation may serve to make some estimates if additional informations
(or simply hints from experiments) are available. This is illustrated below.

Consider the case $p=1$%
\begin{equation}
\frac{2a\alpha }{\sinh \psi _{0}\left( \cosh \psi _{0}-1\right) }=1-\frac{%
a^{2}}{2}\left( \cosh \psi _{0}-1\right) \left( 2\cosh \psi _{0}+1\right)
\label{aalfpsi}
\end{equation}

A \emph{short and dirty} approximation should start by using the suggestion
from results of lucky simulations, where $\psi _{0}$ is few units, and $a$
is of the order $0.1$ on a domain of length $1$ in both $x$ and $y$. The
second derivative of the tangential velocity must be high, shown by the
plots of $v_{\theta }$. This means that it may exist a difference of
magnitude of the terms, with the second term in the right hand side
appearing less important. Therefore we try 
\begin{equation*}
2a\alpha \sim \sinh \psi _{0}\left( \cosh \psi _{0}-1\right)
\end{equation*}
In addition, we can suppose that the exponentials of negative argument are
less important than those of positive argument, and simplify to 
\begin{equation*}
\exp \left( 2\psi _{0}\right) \sim 8a\alpha
\end{equation*}
or 
\begin{equation*}
\psi _{0}\sim \frac{1}{2}\ln \left( \alpha a\right) +1
\end{equation*}

For an order of magnitude we may take 
\begin{equation*}
\alpha \sim \frac{\psi _{0}}{a^{3}}
\end{equation*}
and then 
\begin{eqnarray*}
\psi _{0} &\sim &\frac{1}{2}\ln \left( \frac{\psi _{0}}{a^{2}}\right) +1 \\
&\sim &\frac{1}{2}\ln \psi _{0}-\ln a+1
\end{eqnarray*}
We obtain 
\begin{equation*}
\psi _{0}-1-\frac{1}{2}\ln \psi _{0}\sim -\ln a
\end{equation*}
\begin{eqnarray*}
\frac{1}{a} &\sim &\exp \left( \psi _{0}-1-\frac{1}{2}\ln \psi _{0}\right) \\
&\sim &\frac{1}{\sqrt{\psi _{0}}}\exp \left( \psi _{0}-1\right)
\end{eqnarray*}
or 
\begin{equation*}
a\sim \frac{1}{e}\sqrt{\psi _{0}}\exp \left( -\psi _{0}\right)
\end{equation*}
We can see that the results are consistent, since if we take from numerical
solution 
\begin{equation*}
\psi _{0}\sim 3
\end{equation*}
we obtain from the estimation 
\begin{equation*}
a\sim 0.032
\end{equation*}
which is not far from 
\begin{equation*}
a^{num}\sim 0.04
\end{equation*}
We must remember that the domain of integration is of length $1$ and the
fact that ``the radius of maximum wind'' is so small, $a\sim 0.04$ , means
that high accuracy is needed to describe correctly what happens close to the
center. This is due to the other constraint, that the solution
streamfunction $\psi \left( r\right) $ needs sufficient space to go to the
constant value at ``infinity'' (large $r$). Any restriction of the domain of
integration which would be aimed to the better description of the central
region would require boundary conditions that are unknown.

\bigskip

There is another benefit from these very rough estimations. We can use them
to determine the spatial domain that would be adequate for the search of the
solution, for particular physical situations.

In order to use this rough estimation we must introduce physical units. In
the following all quantities with physical dimensions have an superscript $%
phy$.

In \emph{atmosphere} the distances are measured in $\rho _{g}$%
\begin{equation*}
a=\frac{a^{phy}}{\rho _{g}}
\end{equation*}
and the streamfunction is normalised with 
\begin{equation*}
\psi =\frac{\psi ^{phy}}{\rho _{g}^{2}\left\langle f\right\rangle }
\end{equation*}
where $\left\langle f\right\rangle $ is the Coriolis parameter. This means 
\begin{equation*}
\frac{a^{phy}}{\rho _{g}}\sim \frac{1}{e}\sqrt{\frac{\psi _{0}^{phy}}{\rho
_{g}^{2}\left\langle f\right\rangle }}\exp \left[ -\frac{\psi _{0}^{phy}}{%
\rho _{g}^{2}\left\langle f\right\rangle }\right]
\end{equation*}
The physical parameters are (taken from \cite{HH})

The depth of the atmosphere 
\begin{equation*}
H_{0}=8\times 10^{3}\;\left( m\right)
\end{equation*}

The Coriolis parameter 
\begin{equation*}
\left\langle f\right\rangle =1.6\times 10^{-4}\;\left( s^{-1}\right)
\end{equation*}
From these parameters it results

The Rossby radius (the unit of space) 
\begin{eqnarray*}
\rho _{g} &=&\frac{\left( gH\right) ^{1/2}}{\left\langle f\right\rangle } \\
&=&2\times 10^{6}\,\left( m\right)
\end{eqnarray*}

The unit for the streamfunction is 
\begin{equation*}
\rho _{g}^{2}\left\langle f\right\rangle =6.4\times 10^{8}\;\left(
m^{2}/s\right)
\end{equation*}

The unit for vorticity 
\begin{equation*}
\left\langle f\right\rangle =1.6\times 10^{-4}\;\left( s^{-1}\right)
\end{equation*}

For example, using these parameters, it results that we have integrated on a
spatial domain of length $L$ (in other words: we have imposed that the
streamfunction becomes equal to $\psi _{b}^{\left( 1,2\right) }$ on the
boundaries of a square with side length $L$) 
\begin{eqnarray*}
L &\equiv &x_{\max }-x_{\min }=1 \\
L^{phy} &=&1\times \rho _{g}\sim 2\times 10^{6}\;\left( m\right)
=2000\;\left( km\right) 
\end{eqnarray*}
and the diameter $d$ of the \emph{eye} of the \emph{typhoon} results 
\begin{eqnarray*}
d &=&2\times a=0.08 \\
d^{phy} &\sim &0.08\rho _{g}=128\;\left( km\right) 
\end{eqnarray*}
In the Ref.\cite{mesov} it is reproduced a plot of an observation made on the
profile of the vorticity, in Fig.1a. The plot indicates a maximum value of
about $250\times 10^{-4}\left( s^{-1}\right) $. The vorticity we obtain is
larger (of  the order of $1000 \times 10^{-4}$). 
This shows that the absence of the third dimension
in our model and of the viscous effects have a serious influence on the
physical quantities. They should be somehow accounted for by renormalizing
the two-dimensional model at the initial stage. For example, in the case of 
 the plasma vortex, a change of the space scale results from the presence
of a translational motion combined with the density gradient. This remains to be
studied.

\section{Summary}

We again underline that this equation is very difficult to solve, although
it requires reasonable computer resources. The main problem is the
complexity of the space of solutions and the need to explore carefully much
of this space in order to establish the basins of attraction. We are not
able at this moment to connect in some practically useful way the sharp
transitions between the attractors with the \emph{stability} of the
solutions.

\bigskip

It seems that the solution where the streamfunction $\psi \left( x,y\right) $
is approximately radially symmetric, strongly peaked in origin, is a
significant attractor, at the level of this very sensitive equation. It
presents the particularity that the vorticity is practically zero for almost
all spatial domain and is strongly localised, almost singular, close to the
maximum. The aspect of this solution is very similar to the two-dimensional
image of a \emph{typhoon}.

We have several arguments in favor of the conclusion that our equation (\ref
{eq}) may represent the hydrodynamic part of the atmospheric vortex. We
mention some of them.

\begin{enumerate}
\item  The profile of the magntitude of the tangential velocity, as
represented in Fig.2 of Ref. \cite{cycrev} is very similar to our Fig.\ref
{vthlin_7}. This is also confirmed by similarity with the Fig.1a from Ref( 
\cite{ReMont});

\item  The profile of the vorticity $\omega $ shown in our Fig.\ref{omega_7}
is very similar to Fig.1a from Ref.\cite{mesov};

\item  We note that in \ a series of reported numerical simulations, the
tendency of the fields is to evolve toward profiles that are very close to
those shown in our figures \ref{exp_7c}, \ref{omega_7} and \ref{vthlin_7}.
For example, the Fig.7a and b of Ref.\cite{mesov} show the evolution of the
azimuthal mean of the vorticity and tangential velocity from initial
profiles which correspond to a narrow ring of vorticity to profiles that
show clear ressemblance with our figures \ref{omega_7} and \ref{vthlin_7} or 
\ref{vth_7a}. The same striking evolution to profiles similar to ours
appears in Figs.7 a and b of the same Reference. We have investigated
whether a radially annular profile of vorticity can be a solution of our
equation (\ref{eq}). The result is negative, which may explain why such an
initial profile evolves to either a set of vortices (vortex-crystal) or to a
centrally peaked structure as in Fig.\ref{vth_7a}.

\item  The four vortices represented in Figure 4a of the Ref. \cite{mesov}
as the late stage of the evolution obtained from numerical simulation of
vorticity, is clearly similar to our figure \ref{vel_3}.

\item  We obtain a good consistency between our quantitative results for an
atmospheric vortex (using most elementary input information) and the values
measured or obtained in numerical simulations, at least for some of the
quantities.
\end{enumerate}

A large database on typhoons can be found in \cite{sitety}. The similarity
is striking and it suggests that further work with this equation is worth to
be done.\bigskip

The numerical simulations we have taken as a comparison are very complex. In
general, the physics of the \emph{typhoons} is very complex and includes
hydrodynamics and thermic aspects, with many additional elements:
precipitation, viscosity, etc. In no way we do not claim that this equation
can represent this complexity. It appears however useful as a description of
the regimes where the hydrodynamical processes are dominating and have
reached stationarity.

\bigskip

\textbf{Acknowledgments}. We are very grateful to Professor David Montgomery
for many discussions on a wide spectrum of problems. We thank Dr. L. Weimann
for his kind help on the GIANT code.

This work has been partly supported by a grant from the Japan Society for
the Promotion of Science. The authors are very grateful for this support and
for the hospitality of Professor S.-I. Itoh and of Professor M. Yagi.

\section{Appendix A : The structure of a radial solution near $r=0$ and $%
r=\infty $}

The equation we discuss is 
\begin{equation*}
\Delta \psi +\frac{1}{2p^{2}}\sinh \psi \left( \cosh \psi -p\right) =0
\end{equation*}
Other members of the family of equations (parametrized by solutions of the $%
2D$ Laplace equation) will be examined separately. Their importance stems
from the fact that they can provide, in principle, azimuthal trigonometric
variation, as for example the Larichev-Reznik modon.

\subsection{The behavior near $r=0$}

Close to the origin, in a purely radial form, it is 
\begin{equation*}
\frac{d^{2}\psi }{dr^{2}}+\frac{1}{r}\frac{d\psi }{dr}+\left( \frac{1}{2p^{2}%
}\right) \sinh \psi \left( \cosh \psi -p\right) =0
\end{equation*}
where $r$ is measured in $\rho _{s}$.

We take an expansion with only even powers of $r$ close to the origin 
\begin{equation*}
\psi \sim a_{0}+a_{2}r^{2}+a_{4}r^{4}+a_{6}r^{6}+...
\end{equation*}
Then, for small $r$; 
\begin{eqnarray*}
\frac{d\psi }{dr} &=&2a_{2}r+4a_{4}r^{3}+6a_{6}r^{5}... \\
\frac{1}{r}\frac{d\psi }{dr} &=&2a_{2}+4a_{4}r^{2}+6a_{6}r^{4}...
\end{eqnarray*}
\begin{equation*}
\frac{d^{2}\psi }{dr^{2}}=2a_{2}+12a_{4}r^{2}+30a_{6}r^{4}+...
\end{equation*}
\begin{eqnarray*}
&&\sinh \left( a_{0}+a_{2}r^{2}+a_{4}r^{4}+...\right) \\
&=&\sinh a_{0} \\
&&+\left( a_{2}r^{2}+a_{4}r^{4}+...\right) \cosh a_{0} \\
&&+\frac{1}{2}\left( a_{2}^{2}r^{4}+...\right) \sinh a_{0}+...
\end{eqnarray*}
\begin{eqnarray*}
&&\cosh \left( a_{0}+a_{2}r^{2}+a_{4}r^{4}+...\right) \\
&=&\cosh a_{0} \\
&&+\left( a_{2}r^{2}+a_{4}r^{4}+...\right) \sinh a_{0} \\
&&+\frac{1}{2}\left( a_{2}^{2}r^{4}+...\right) \cosh a_{0}+...
\end{eqnarray*}
Introducing the notations 
\begin{eqnarray*}
U &\equiv &a_{2}r^{2}+a_{4}r^{4}+... \\
V &\equiv &\frac{1}{2}\left( a_{2}^{2}r^{4}+...\right)
\end{eqnarray*}
\begin{eqnarray*}
&&\sinh \psi \left( \cosh \psi -p\right) \\
&=&\left( \sinh a_{0}+U\cosh a_{0}+V\sinh a_{0}\right) \\
&&\times \left( \cosh a_{0}-p+U\sinh a_{0}+V\cosh a_{0}\right) \\
&=&\sinh a_{0}\left( \cosh a_{0}-p\right) \\
&&+U\left( \sinh ^{2}a_{0}+\cosh ^{2}a_{0}-p\cosh a_{0}\right) \\
&&+V\left( 2\cosh a_{0}\sinh a_{0}-p\sinh a_{0}\right) \\
&&+U^{2}\left( \sinh a_{0}\cosh a_{0}\right) \\
&&+V^{2}\left( \sinh a_{0}\cosh a_{0}\right) \\
&&+UV\left( \cosh ^{2}a_{0}+\sinh ^{2}a_{0}\right) \\
&&+...
\end{eqnarray*}
We collect the various degrees of $r^{\alpha }$%
\begin{equation*}
q_{0}+q_{2}r^{2}+q_{4}r^{4}+...
\end{equation*}
\begin{equation*}
q_{0}=\sinh a_{0}\left( \cosh a_{0}-p\right)
\end{equation*}
\begin{equation*}
q_{2}=a_{2}\left( \sinh ^{2}a_{0}+\cosh ^{2}a_{0}-p\cosh a_{0}\right)
\end{equation*}
\begin{eqnarray*}
q_{4} &=&a_{4}\left( \sinh ^{2}a_{0}+\cosh ^{2}a_{0}-p\cosh a_{0}\right) \\
&&+\frac{1}{2}a_{2}^{2}\left( 2\cosh a_{0}\sinh a_{0}-p\sinh a_{0}\right) \\
&&+a_{2}^{2}\left( \sinh a_{0}\cosh a_{0}\right)
\end{eqnarray*}
Returning to the differential operator 
\begin{eqnarray*}
&&\frac{d^{2}\psi }{dr^{2}}+\frac{1}{r}\frac{d\psi }{dr} \\
&=&2a_{2}+12a_{4}r^{2}+30a_{6}r^{4}+... \\
&&+2a_{2}+4a_{4}r^{2}+6a_{6}r^{4}... \\
&=&4a_{2}+16a_{4}r^{2}+36a_{6}r^{4}+...
\end{eqnarray*}
We now identify the expressions corresponding to the same degrees of $r$, 
\begin{eqnarray*}
&&4a_{2}+16a_{4}r^{2}+36a_{6}r^{4}+... \\
&&+\left( \frac{1}{2p^{2}}\right) \left(
q_{0}+q_{2}r^{2}+q_{4}r^{4}+...\right) \\
&=&0
\end{eqnarray*}
with the equalities 
\begin{equation*}
4a_{2}+\left( \frac{1}{2p^{2}}\right) q_{0}=0
\end{equation*}
\begin{equation*}
16a_{4}+\left( \frac{1}{2p^{2}}\right) q_{2}=0
\end{equation*}
\begin{equation*}
36a_{6}+\left( \frac{1}{2p^{2}}\right) q_{4}=0
\end{equation*}
The equations from which we derive the coefficients of the expansion become 
\begin{equation*}
4a_{2}+\left( \frac{1}{2p^{2}}\right) \sinh a_{0}\left( \cosh a_{0}-p\right)
=0
\end{equation*}
\begin{equation*}
16a_{4}+\left( \frac{1}{2p^{2}}\right) a_{2}\left( \sinh ^{2}a_{0}+\cosh
^{2}a_{0}-p\cosh a_{0}\right) =0
\end{equation*}
\begin{eqnarray*}
&&36a_{6}+\left( \frac{1}{2p^{2}}\right) \left[ a_{4}\left( \sinh
^{2}a_{0}+\cosh ^{2}a_{0}-p\cosh a_{0}\right) \right. \\
&&+\frac{1}{2}a_{2}^{2}\left( 2\cosh a_{0}\sinh a_{0}-p\sinh a_{0}\right) \\
&&\left. +a_{2}^{2}\left( \sinh a_{0}\cosh a_{0}\right) \right] \\
&=&0
\end{eqnarray*}

We see that if we take 
\begin{equation*}
a_{0}=0
\end{equation*}
then this will vanish all the other coefficients 
\begin{eqnarray*}
a_{2} &=&0 \\
a_{4} &=&0 \\
a_{6} &=&0,...
\end{eqnarray*}

\begin{figure}[tbph]
\centerline{\includegraphics[height=5cm]{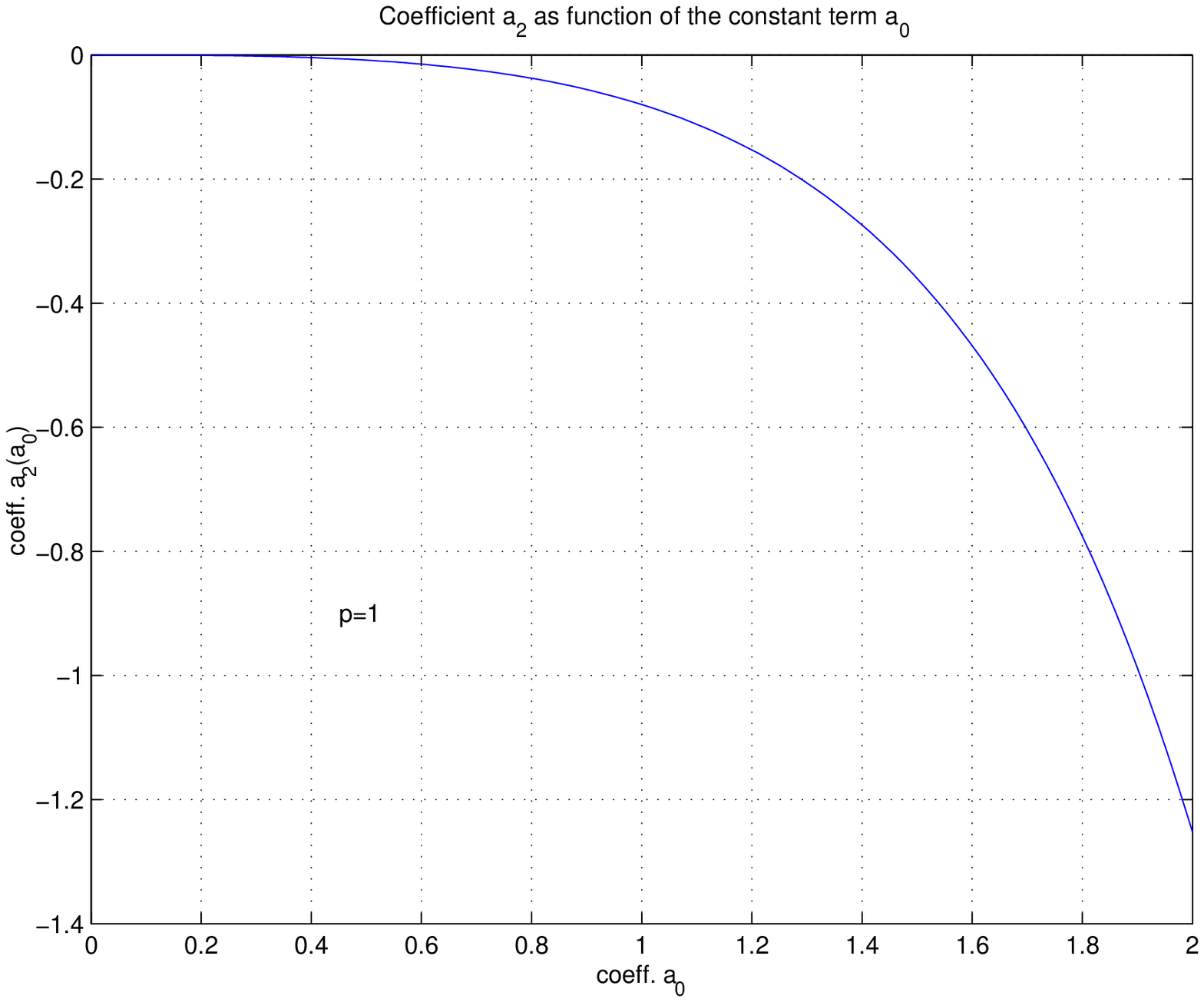}}
\caption{Coefficient $a_{2}$ for $p=1$.}
\label{a2_1}
\end{figure}

\begin{figure}[tbph]
\centerline{\includegraphics[height=5cm]{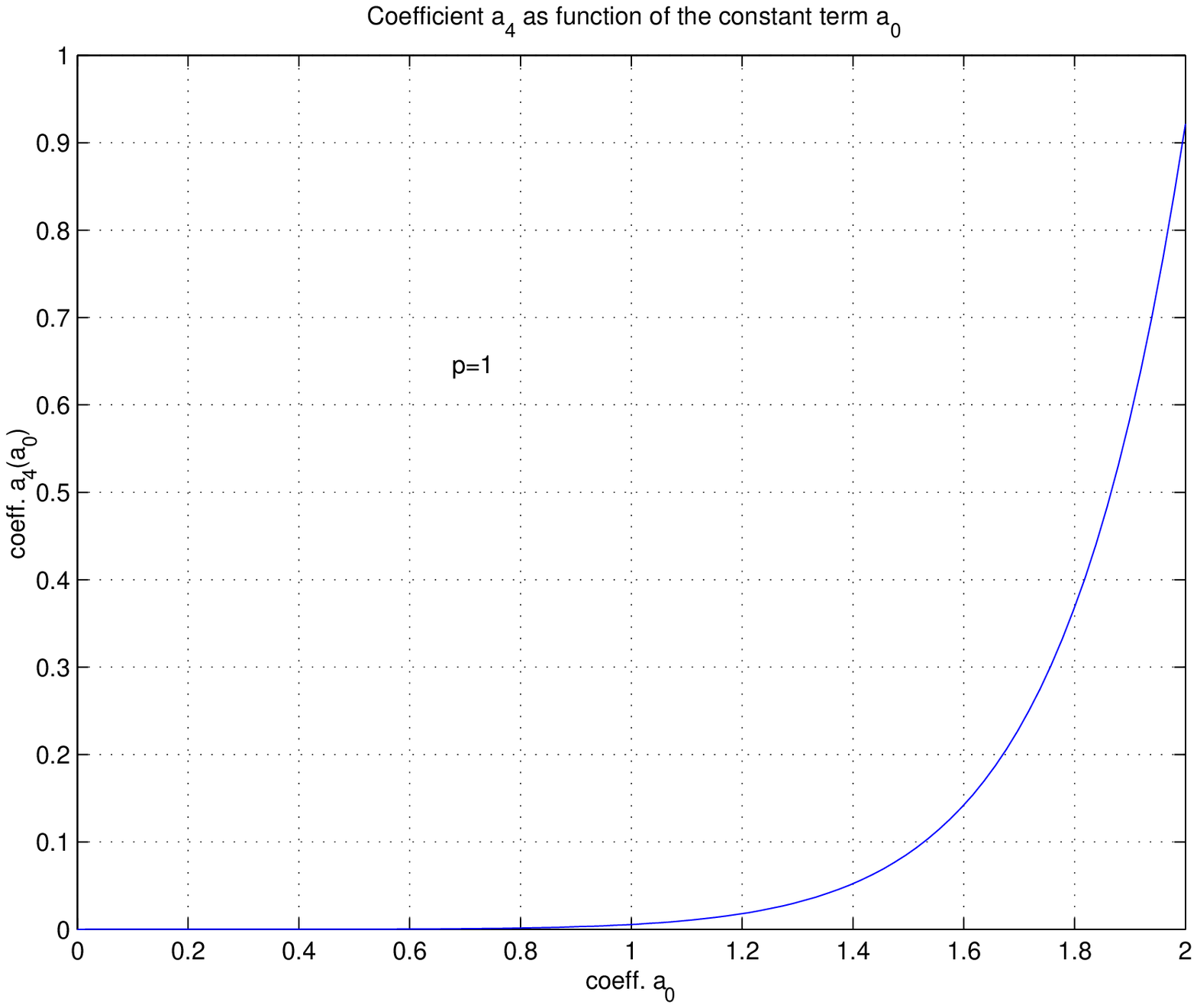}}
\caption{Coefficient $a_{4}$ for $p=1$.}
\label{a4_1}
\end{figure}

\begin{figure}[tbph]
\centerline{\includegraphics[height=5cm]{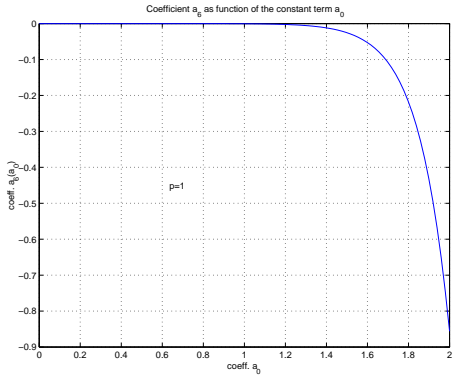}}
\caption{Coefficient $a_{6}$ for $p=1$.}
\label{a6_1}
\end{figure}

Consider the value of the constant 
\begin{equation*}
p=1
\end{equation*}
and we choose the main coefficient of the expansion close to $r=0$ to be 
\begin{equation*}
a_{0}=1
\end{equation*}
Then 
\begin{eqnarray*}
a_{2} &=&-0.0798 \\
a_{4} &=&0.0055 \\
a_{6} &=&-0.000439
\end{eqnarray*}
But the coefficients, as shown in the Figures, are very rapidly growing in
absolute value.

We conclude that any attempt to identify the solution starting from few
terms expansion around $r=0$ will be imprecise.

\subsection{The behavior at infinity}

At $r\rightarrow \infty $ we expect that the function approaches zero in the
case where $p=1$ or approaches one of the roots of the equation 
\begin{equation}
\cosh \psi -p=0  \label{psibeq}
\end{equation}
for $p>1$. The case where $\psi \rightarrow 0$ will be treated below. We
note, for the case $p>1$ that the solutions of the Eq.(\ref{psibeq}) are 
\begin{eqnarray*}
\psi _{b}^{\left( 1\right) } &=&\ln \left( p+\sqrt{p^{2}-1}\right) \\
\psi _{b}^{\left( 2\right) } &=&\ln \left( p-\sqrt{p^{2}-1}\right)
\end{eqnarray*}

\subsubsection{The case $p=1$}

This requires that $\psi \rightarrow 0$ at $r\rightarrow \infty $.

Change the variable 
\begin{equation*}
r\rightarrow \frac{1}{x}
\end{equation*}
\begin{equation*}
\frac{d}{dr}=\frac{dx}{dr}\frac{d}{dx}=-\frac{1}{r^{2}}\frac{d}{dx}=-x^{2}%
\frac{d}{dx}
\end{equation*}
\begin{eqnarray*}
\frac{d^{2}}{dr^{2}} &=&\frac{d}{dr}\left( \frac{d}{dr}\right) =-x^{2}\frac{d%
}{dx}\left( -x^{2}\frac{d}{dx}\right) \\
&=&-x^{2}\left( -2x\frac{d}{dx}-x^{2}\frac{d^{2}}{dx^{2}}\right) \\
&=&2x^{3}\frac{d}{dx}+x^{4}\frac{d^{2}}{dx^{2}}
\end{eqnarray*}
The function 
\begin{equation*}
\psi \rightarrow 0
\end{equation*}
\begin{eqnarray*}
\left( \frac{1}{2p^{2}}\right) \sinh \psi \left( \cosh \psi -p\right)
&\rightarrow &\left( \frac{1}{2p^{2}}\right) \left( \psi -\frac{\psi ^{3}}{6}%
\right) \left( 1-p-\frac{\psi ^{2}}{2}\right) \\
&=&\frac{1-p}{2p^{2}}\psi \\
&&+\frac{1}{2p^{2}}\left( -\frac{1}{2}-\frac{1-p}{6}\right) \psi ^{3}+...
\end{eqnarray*}
For 
\begin{eqnarray*}
p &=&1 \\
\left( \frac{1}{2p^{2}}\right) \sinh \psi \left( \cosh \psi -p\right)
&\rightarrow &-\frac{1}{4}\psi ^{3}
\end{eqnarray*}
Then the equation becomes 
\begin{eqnarray*}
&&\left( 2x^{3}\frac{d}{dx}+x^{4}\frac{d^{2}}{dx^{2}}\right) \psi \\
&&+\left( -x^{2}\frac{d}{dx}\right) \psi \\
&&+\frac{1-p}{2p^{2}}\psi +\frac{1}{2p^{2}}\left( -\frac{1}{2}-\frac{1-p}{6}%
\right) \psi ^{3} \\
&=&0
\end{eqnarray*}
This can be approximated at 
\begin{equation*}
x\rightarrow 0
\end{equation*}
\begin{equation*}
-x^{2}\frac{d\psi }{dx}=\alpha \psi +\beta \psi ^{3}
\end{equation*}
or 
\begin{equation*}
\frac{d\psi }{\alpha \psi +\beta \psi ^{3}}=-\frac{dx}{x^{2}}=d\left( \frac{1%
}{x}\right) =dr
\end{equation*}
For 
\begin{equation*}
p=1
\end{equation*}
\begin{eqnarray*}
\alpha &=&0 \\
\beta &=&-\frac{1}{4}
\end{eqnarray*}
then 
\begin{equation*}
\left( -4\right) \frac{d\psi }{\psi ^{3}}=dr
\end{equation*}
\begin{equation*}
\psi \sim \sqrt{\frac{2}{r}}
\end{equation*}
We note however that in this case the vorticity is 
\begin{eqnarray*}
\omega &=&\Delta \psi \\
&\sim &r^{-5/2}
\end{eqnarray*}
We would like to have a vanishing vorticity at infinity with a faster decay.

The above calculations seem to suggest that for purely radial structure we
need to consider the differential equation which is derived for a different
choice of the Laplacean equation, as it is explained in the main text.

\subsubsection{The case $p>1$}

One possibility, for 
\begin{equation*}
p>1
\end{equation*}
\begin{equation*}
\alpha \equiv \frac{1-p}{2p^{2}}<0
\end{equation*}
\begin{equation*}
\psi \sim \exp \left( -\left| \alpha \right| r\right)
\end{equation*}
This gives 
\begin{eqnarray*}
\omega &=&\Delta \psi \\
&\sim &\left( -\left| \alpha \right| \right) \frac{\exp \left( -\left|
\alpha \right| r\right) }{r}+\alpha ^{2}\exp \left( -\left| \alpha \right|
r\right)
\end{eqnarray*}
with a fast decay. This situation is worth to be examined numerically.

\section{Appendix B : various forms of the initial conditions}

\subsection{The ring-type}

The initial form of the function has the form 
\begin{equation*}
\psi _{0}=A\exp \left( -sr^{2}\right) \left[ 1-\kappa \exp \left(
-qr^{4}\right) \right]
\end{equation*}
We look for the maximum 
\begin{eqnarray*}
\frac{d\psi _{0}}{dr} &=&\left( -2sr\right) \exp \left( -sr^{2}\right) \left[
1-\kappa \exp \left( -qr^{4}\right) \right] \\
&&+\exp \left( -sr^{2}\right) \left( 4qr^{3}\right) \kappa \exp \left(
-qr^{4}\right) \\
&=&0
\end{eqnarray*}
and we take the maximum to be placed at 
\begin{equation*}
r=a
\end{equation*}
which is considered to approximate the center line of the ring. The equation
becomes 
\begin{equation*}
\left( 2s\kappa +4\kappa qa^{2}\right) \exp \left( -qa^{4}\right) =2s
\end{equation*}
\begin{equation*}
\kappa \exp \left( -qa^{4}\right) =\frac{1}{1+2a^{2}\left( q/s\right) }
\end{equation*}
The other condition is that the maximum of the function $\psi _{0}$ at $r=a$
equals a prescribed value, 
\begin{eqnarray*}
\psi _{0}\left( r=a\right) &=&\psi _{c} \\
A\exp \left( -sa^{2}\right) \left[ 1-\kappa \exp \left( -qa^{4}\right) %
\right] &=&\psi _{c}
\end{eqnarray*}
The initial condition is introduced in the following way. We take $q$, $a$, $%
\kappa $ and $\psi _{c}$ as input parameters and determine the other two, $s$
and $A$ from the equations 
\begin{equation*}
s=\frac{2a^{2}q}{\kappa \exp \left( -qa^{4}\right) -1}
\end{equation*}
\begin{equation*}
A=\frac{\psi _{c}}{\exp \left( -sa^{2}\right) \left[ 1-\kappa \exp \left(
-qa^{4}\right) \right] }
\end{equation*}

Now the initial function will be 
\begin{eqnarray*}
\psi _{initial}\left( r\right) &=&\psi _{0}+\psi _{b}^{\left( 1,2\right) } \\
&=&\psi _{b}^{\left( 1,2\right) }+ \\
&&+A\exp \left( -sr^{2}\right) \left[ 1-\kappa \exp \left( -qr^{4}\right) %
\right]
\end{eqnarray*}
\emph{i.e.} the function just determined is placed on the constant
background of the value at the boundary, calculated form the condition that
the vorticity is zero at infinity.

This class of initial functions is characterised by an annular shape, with
exponential decay for $r\rightarrow \infty $, with a minimum in the region
around $r=0$ of depth that can be fixed by varying $\kappa $. For $\kappa =1$
the function is zero on the symmetry axis and rises slowly (due to $r^{4}$)
toward the maximum at $r=a$.

In order to narrow the space of parameters we require the approximative
equality between the vorticity amplitude at the ring with the nonlinear term 
\begin{eqnarray*}
\omega &\sim &-\frac{2}{\delta ^{2}}\psi _{c} \\
&\sim &-\frac{1}{2p^{2}}\sinh \left( \psi _{c}+\psi _{b}^{\left( 1,2\right)
}\right) \left[ \cosh \left( \psi _{c}+\psi _{b}^{\left( 1,2\right) }\right)
-p\right]
\end{eqnarray*}
(Here $\delta $ is the width of the ring shape). These two quantities are
compared in graphical plot for a range of values of the parameter $\psi _{c}$%
, using a Matlab script. This is far from an exact procedure but helps to
generate reasonable ranges for the input parameters.

\bigskip

The conclusion after many trials using this procedure and its initial
function forms can be described as follows.

In most of the cases the central region is corrected and shifted to a
maximum. In the cases $p=1$ the central region which is started with a
deppressed level is rised and a strong peaked form is generated, as in the
cases where the initialization consists of a maximum on center (for example
a Gaussian form). For $p>1$ the run evolves in some cases to the formation
of separate maxima placed symmetrically on a ring, having sharp maxima. The
central region is decreased in amplitude to a somehow flat region. The
region outside the ring is evolving to a state which corresponds with very
good precision, to 
\begin{equation*}
\omega \sim 0
\end{equation*}
on the rest of the domain to the periphery.

\subsection{Flat central region for $\protect\psi \left( r\right) $}

We take the central region 
\begin{equation*}
0<r<r_{flat}
\end{equation*}
with a fixed, constant value 
\begin{equation*}
\psi \left( r\right) =\psi _{c}
\end{equation*}
where $\psi _{c}$ is one of the roots of the equation $\cosh \psi -p=0$. At
the edge we take another fixed value, 
\begin{equation*}
\psi =\psi _{b}
\end{equation*}
with $\psi _{b}$ the other, smaller root of the equation.

In between, we take 
\begin{equation*}
\psi \left( r\right) =\psi _{1}-A\ln \left( r\right)
\end{equation*}
\begin{equation*}
A=\frac{\psi _{c}-\psi _{b}}{\ln \left( r_{flat}/r_{c}\right) }
\end{equation*}
\begin{equation*}
\psi _{1}=\psi _{c}+A\ln \left( r_{flat}\right)
\end{equation*}
The value $r_{c}$ is 
\begin{equation*}
r_{flat}<r_{c}<0.5
\end{equation*}
represents the value where where we stop the decay of the function with
logarithm profile and put $\psi =\psi _{b}$. This is 
\begin{eqnarray*}
r_{flat} &=&0.1 \\
r_{c} &=&0.35\cdots 0.45
\end{eqnarray*}
The parameter $p=1.3$.

The result of these calculations is as follows. For small mesh, the
evolution is clearly toward the suppression of the smoothly decaying part,
letting a sort of cylinder in the center, with radius $r_{flat}$, with the
high value equal to $\psi _{c}$ and the rest seems to go progressively to $%
\psi =\psi _{b}$. The vorticity is singular, around $r=r_{flat}$. The
vorticity is positive and negative, with high values, singular in a narrow
ring.

For this cylindrical-rod profile of the streamfunction $\psi \left( r\right) 
$, the velocity is very localised, as a very narrow ring, all its values are
positive. The velocity grows from zero, keeps always the same direction on $%
\theta $ and then decays to zero value, after the width of the ring. The
vorticity is also sharply limitted here, but it has positive and negative
values on interior half of the ring and respectively on the exterior half of
the ring.

The same shrinking to the cylindrical column happens when we take the
maximum of $\psi $ (in the central flat region) as 
\begin{equation*}
\psi \left( r\right) =0
\end{equation*}
which is the other possibility that the equation is verified for constant
value of $\psi $.

\end{document}